\documentclass[preprint,flushrt]{aastex}
\usepackage{natbib}
\usepackage{lineno}
\linenumbers

\shorttitle{LBAS SED}
\shortauthors{Abdo et al.}

\title{Spectral Properties of Bright Fermi-detected Blazars in the Gamma-ray 
Band }

\author{
A.~A.~Abdo\altaffilmark{1,2}, 
M.~Ackermann\altaffilmark{3}, 
M.~Ajello\altaffilmark{3}, 
W.~B.~Atwood\altaffilmark{4}, 
M.~Axelsson\altaffilmark{5,6}, 
L.~Baldini\altaffilmark{7}, 
J.~Ballet\altaffilmark{8}, 
G.~Barbiellini\altaffilmark{9,10}, 
D.~Bastieri\altaffilmark{11,12}, 
K.~Bechtol\altaffilmark{3}
% ...
}
\altaffiltext{1}{Space Science Division, Naval Research Laboratory, Washington, DC 20375, USA}
\altaffiltext{2}{National Research Council Research Associate, National Academy of Sciences, Washington, DC 20001, USA}
\altaffiltext{3}{W. W. Hansen Experimental Physics Laboratory, Kavli Institute for Particle Astrophysics and Cosmology, Department of Physics and SLAC National Accelerator Laboratory, Stanford University, Stanford, CA 94305, USA}
\altaffiltext{4}{Santa Cruz Institute for Particle Physics, Department of Physics and Department of Astronomy and Astrophysics, University of California at Santa Cruz, Santa Cruz, CA 95064, USA}
\altaffiltext{5}{Department of Astronomy, Stockholm University, SE-106 91 Stockholm, Sweden}
\altaffiltext{6}{The Oskar Klein Centre for Cosmoparticle Physics, AlbaNova, SE-106 91 Stockholm, Sweden}
\altaffiltext{7}{Istituto Nazionale di Fisica Nucleare, Sezione di Pisa, I-56127 Pisa, Italy}
\altaffiltext{8}{Laboratoire AIM, CEA-IRFU/CNRS/Universit\'e Paris Diderot, Service d'Astrophysique, CEA Saclay, 91191 Gif sur Yvette, France}
\altaffiltext{9}{Istituto Nazionale di Fisica Nucleare, Sezione di Trieste, I-34127 Trieste, Italy}
\altaffiltext{10}{Dipartimento di Fisica, Universit\`a di Trieste, I-34127 Trieste, Italy}
\altaffiltext{11}{Istituto Nazionale di Fisica Nucleare, Sezione di Padova, I-35131 Padova, Italy}
\altaffiltext{12}{Dipartimento di Fisica ``G. Galilei", Universit\`a di Padova, I-35131 Padova, Italy}
% ...
\begin{abstract}
The gamma-ray energy spectra of bright blazars of the LAT Bright AGN Sample (LBAS) are investigated using {\sl Fermi}-LAT data. Spectral properties (hardness, curvature and variability) established using a data set accumulated over 6 months of operation are presented and discussed for different blazar classes and subclasses: Flat Spectrum Radio Quasars (FSRQs), Low-synchrotron peaked BLLacs (LSP-BLLacs), Intermediate-synchrotron peaked BLLacs (ISP-BLLacs) and High-synchrotron peaked BLLacs (HSP-BLLacs). The distribution of photon index ($\Gamma$, obtained from a power-law fit above 100 MeV) is found to correlate strongly with blazar subclass. The change in spectral index from that averaged over the six month observing period is $<$ 0.2-0.3 when the flux varies by about an order of magnitude, with a tendency toward harder spectra when the flux is brighter for FSRQs and LSP-BLLacs. A strong departure from a single power-law spectrum appears to be a common feature for FSRQs. This feature is also present for some high-luminosity LSP-BLLacs, and a small number of ISP-BLLacs. It is absent in all LBAS HSP-BLLacs.  For 3C\,454.3 and AO\,0235+164, the two brightest FSRQ source and LSP-BLLac source respectively, a broken power law gives the most acceptable of power law, broken power law, and curved forms. The consequences of these findings are discussed.

\end{abstract}


\begin{document}

\section{Introduction}

Launched into a low-Earth orbit on June 11, 2008, the {\it Fermi Gamma-Ray Space Telescope} continues providing excellent gamma-ray data for celestial sources.  With significant improvement of sensitivity and bandpass over its predecessors \citep{LATpaper}, the main instrument on {\it Fermi} - the Large Area Telescope, or LAT - enables detailed studies of time-resolved broad-band gamma-ray spectra of a broad range of sources, including active galaxies.  As discovered by EGRET on the Compton Observatory \citep{Har92a, Fic94}, active galactic nuclei (AGN) showing strong gamma-ray emission are associated with relativistic jets, whose presence was independently inferred from morphological and variability studies in other bands.  The spectra of such objects in all observable bands are well-described by broad power-law or curved distributions, indicating non-thermal emission mechanisms \citep{Boe07}.  Very generally, the overall broad-band spectral distributions of such jet-dominated AGN, often called blazars have a two-humped shape, with the low energy (IR-UV) hump attributed to synchrotron emission of energetic electrons radiating in magnetic field and the high energy hump due to inverse Compton scattering by the same electrons \citep{Ghi89, Der93, Sik94}.

The first list of such AGN detected by the {\it Fermi}-LAT, the LAT Bright AGN Sample (LBAS) \citep{LBAS} includes bright, high-galactic latitude ($|b|> 10^{\rm o}$) AGNs detected by the Fermi-LAT with high significance (Test Statistic TS$>100$) during the first three months of scientific operation.  This sample comprises 58 Flat Spectrum Radio Quasars (FSRQs), 42 BLLac-type objects (BLLacs), two radio galaxies and four 
quasars of unknown type.  This somewhat conventional classification was based on the observed optical emission line equivalent widths and the Ca II break ratio \citep[e.g.,][]{mar96}.  Following the models used 
to describe the gamma-ray spectra  obtained with previous gamma-ray observatories \citep[e.g.,][]{mat96}, the early analysis reported  in \citep{LBAS} was carried out by fitting the gamma-ray spectra at energies above 200 MeV using a simple power law (PL) model. This analysis revealed  a fairly distinct spectral separation between FSRQs and BLLacs,  with FSRQs having significantly softer spectra.  The boundary photon index between the two classes was found to be  $\Gamma \simeq$2.2.  It has been suggested \citep{Ghi09} that this separation results from different radiation cooling suffered by the electrons due to  distinct accretion regimes in the two blazar classes. 

While adopting such a simple spectral model was sufficient  to investigate the source spectral hardness distribution,  a PL model was clearly not the most appropriate choice  for some bright sources which exhibited evident breaks or curvatures  in their spectra.  The departure of the functional form from a PL was  investigated in some detail for the bright quasar  3C\,454.3 \citep{LAT3C454.3},  which underwent strong activity in the summer of 2008. The change of  photon index $\Delta \Gamma$ was observed to be 1.2$\pm$0.2, i.e. greater than  the value of 0.5  expected from incomplete cooling of the emitting electrons. The  observed break around 2.2 GeV was ascribed to mirroring a similar  feature in the underlying emitting electron energy distribution;  the Klein-Nishina effect was not ruled out, though the importance of photon-photon pair production requires the gamma-ray emission region to be close to the  supermassive black hole. Clearly, understanding the details of the  spectral break is important for understanding the structure and location of the dissipation region of jets in active galaxies.

The data first obtained with the EGRET instrument, now refined with Fermi, imply  that the high Galactic latitude sky emits quasi-diffuse, uniform gamma-ray background \citep{Sre98,  Str04, Ack09}. Its isotropy points to its extragalactic nature after subtraction of a quasi-isotropic gamma-ray emission component from cosmic ray electrons in an extended  galactic halo. Most models  account for at least a part of this background as originating from a large number of unresolved point sources, presumably jet-dominated AGN. Here, the comparison of spectral properties of various classes of AGN, integrated over their space density and luminosity,  against the integral measurement of the unresolved component should provide  additional clues regarding their contribution to the  extragalactic diffuse background: can they make up the entire background, is another class of sources, e. g. star-forming galaxies \citep{Fie09} or is an additional truly diffuse component required?   The issue is further complicated by the apparently different spectral  forms of the luminous AGN associated with quasars as compared to the  lineless BL Lac objects, as already hinted in \cite{LBAS}.   It is thus important to determine whether the  spectral feature seen in 3C\,454.3 is common in all blazars and also whether it is connected to other blazar properties. 

Here we report on the detailed spectral analysis of bright LBAS sources using data accumulated over the first 6 months of the  {\it Fermi}-LAT all-sky survey.  In Section 2, we present the observations  with the {\it Fermi}-LAT; Section 3 briefly discusses the classification scheme  used in this paper. Section 4 contains the results regarding the photon index  and observed deviations from a pure PL.   The results and their consequences are discussed in Section 5.   

\section{Observations with the Large Area Telescope}

The {\it Fermi}-LAT is a pair-conversion gamma-ray telescope sensitive to photon  energies greater than 20 MeV. It is made of a tracker (composed of two sections,  front and back, with different localization capabilities), a calorimeter, and an  anticoincidence system to reject the charged-particle background. The LAT has a  large peak effective area ($\sim 8000$ cm$^2$ for 1 GeV on-axis photons in the  event class ``diffuse'' considered here), viewing $\approx 2.4$ sr of the full  sky with an angular resolution (68\% containment angle) better than $ \approx  1^\circ$ at $E = 1$ GeV \citep{LATpaper}.

The data were collected from 4 Aug. 2008 to 1 Feb. 2009 in survey mode.  To  minimize systematics, only photons with energies greater than 100 MeV were  considered in this analysis. In order to avoid contamination from Earth limb  gamma-rays, a selection on the zenith angle, $<105^{\circ}$, was applied. The  exposure was constant within 20\% for all sources and amounted to about  1.5$\times$10$^6$ m$^2$s at 1 GeV. 

This analysis was performed with the standard analysis tool {\it gtlike}, part  of the Fermi-LAT ScienceTools software package (version v9r12). The first set of  instrument response functions (IRFs) tuned with the flight data,  P6\_V3\_DIFFUSE, was used. In contrast to the preflight IRFs, these IRFs take  into account corrections for pile-up effects. This correction being higher for  lower energy photons, the measured photon index of a given source is about 0.1  higher (i.e. the spectrum is softer) with this IRF set as compared to the  P6\_V1\_DIFFUSE one used previously in \cite{LBAS}. Photons were selected in  circular regions of interest (ROI), 7$^\circ$ in radius, centered at the  positions of the sources of interest.  The isotropic background (the sum of  residual instrumental background and extragalactic diffuse gamma-ray background)  was modeled with a simple power-law. The GALPROP model \citep{Str04, Str04b},  version ``gll\_iem\_v01.fit'', was used for the galactic diffuse emission, with  both flux and spectral photon index left free in the fit.  All point sources  with TS$>$25 in the 6-month source list, lying within the ROI and a surrounding  5$^\circ$-wide annulus, were modeled in the fit with single power-law  distributions.  Different analyses were performed by fitting the spectra with  various models over the whole energy range covered by the LAT above 100 MeV, or  with a PL model over equispaced logarithmic energy bins (where the spectral  index was kept constant and equal to the value fitted over the whole range). In  the case of fits with broken power law (BPL) models, the break energy  (E$_{Break}$) bounding the ranges where different photon indices ($\Gamma_1$ and  $\Gamma_2$) apply, could not be obtained directly from the fit for most sources  because of convergence problem due the non-smooth character of the BPL function at the break energy. It was computed from  a loglikelihood profile fitting procedure, with statistical uncertainties  corresponding to a difference of -2$\Delta L$=1 in the loglikelihood (L) with  respect to its minimum. We refer the reader to ref. \citep{Dag04} regarding limitations with the use of asymmetric uncertainties.

The estimated systematic uncertainty on the flux is 10\% at 100 MeV, 5\% at 500  MeV and 20\% at 10 GeV. The energy resolution is better than 10\% over the range  of measured E$_{Break}$.

\section{Classification}

We employ the conventional definition of BL Lac objects outlined in  \cite{sto91, up95, mar96} in which the equivalent width of the strongest optical  emission line is $<$ 5 \AA \- and the optical spectrum shows a Ca II H/K break  ratio C $<$ 0.4.   BLLac sources were assigned to different subclasses (LSP-BLLacs, ISP-BLLacs and HSP-BLLacs  standing for Low-, Intermediate-, and High-synchrotron peaked BLLacs respectively)  according to the position of their synchrotron peak, established from radio,  optical, UV and X-ray data: $\nu_{peak} < 10^{14}$ Hz for LSP-BLLacs, $10^{14}$Hz$<  \nu_{peak} < 10^{15}$Hz for ISP-BLLacs and $\nu_{peak}>10^{15}$Hz for HSP-BLLacs  \citep{SEDpaper}.  Contemporaneous Swift data were used for a subset of 46 LBAS  sources and archival ones for the others, as described in \cite{SEDpaper}.

\section{Results}

\subsection{Photon index distributions}
 
\begin{figure} \epsscale{0.6} 
\plotone{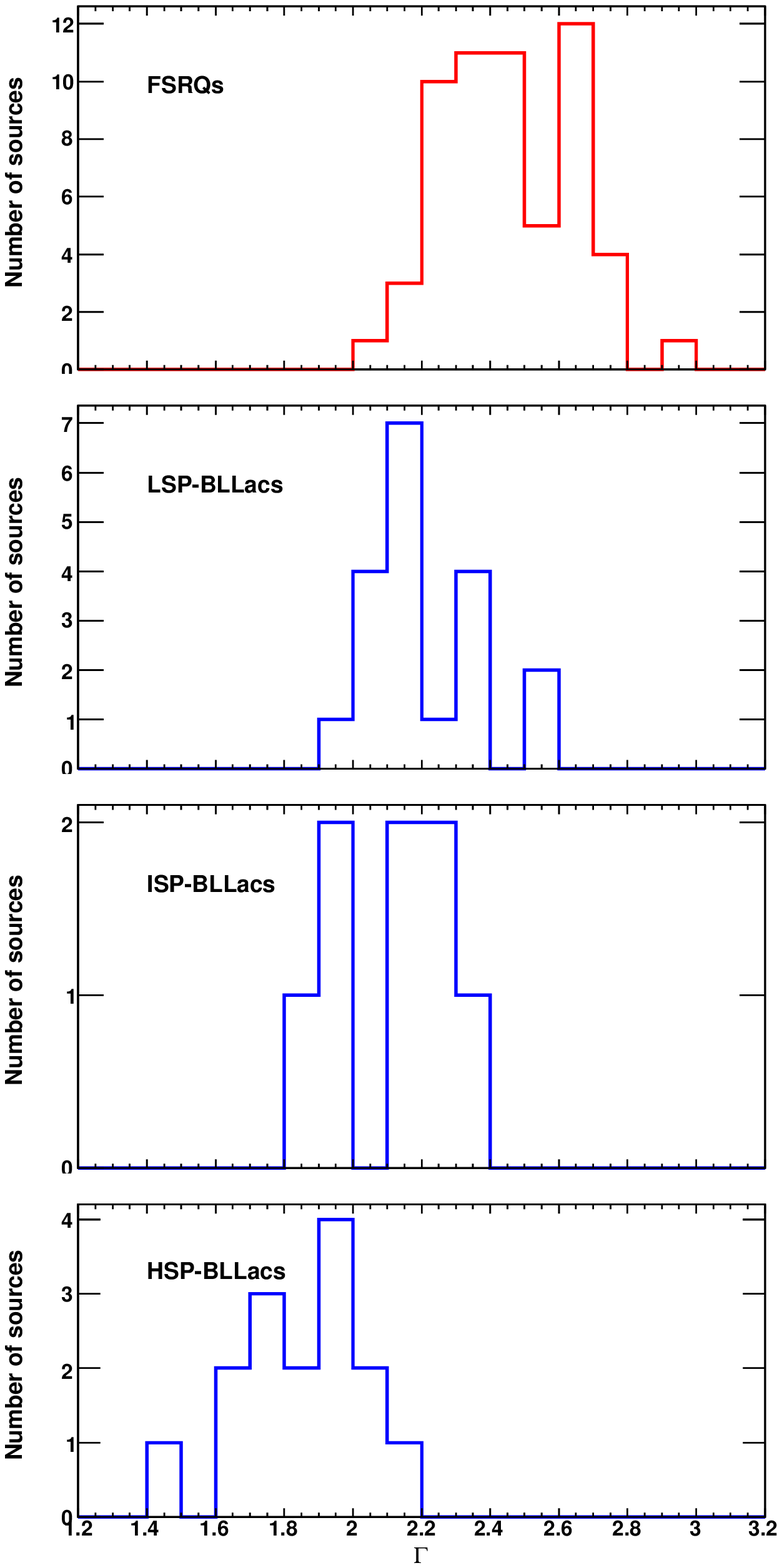} 
\caption{Gamma-ray photon index distributions for the four blazar subclasses.} 
\label{fig:index}
\end{figure}

Although some spectra display significant curvatures, the photon index obtained  by fitting single power-paw models over the whole LAT energy range provides a  convenient means to study the spectral hardness.  Fig. \ref{fig:index} displays the distributions of the resulting photon index  for the four different subclasses. The remarkable separation between FRSQs and  BLLacs already found in \cite{LBAS} is of course still observed for spectra  averaged over a 6-month (instead of 3-month) time span. Likewise, different  BLLac subclasses are associated with distinct photon index distributions in the  LAT range. The distributions have $(\mathrm{mean}, \mathrm{rms})$=(2.46, 0.18)  for FSRQs, (2.21, 0.16) for LSP-BLLacs, (2.13, 0.17) for ISP-BLLacs and (1.86, 0.17) for  HSP-BLLacs. For comparison, the distributions given in \cite{LBAS} had  $(\mathrm{mean}, \mathrm{rms})$= (2.40, 0.17), (1.99, 0.22) for FSRQs and BL  Lacs respectively. It must be kept in mind that the 6-month distributions have  been obtained with an improved (more realistic) IRF set leading to a softer  measured spectrum ($\Gamma$ higher by $\simeq$0.1 unit). Interestingly, the  largest difference (increase) in photon index between the 3-month and the 6-month data set is obtained for BL Lacertae (from $\Gamma$=2.24$\pm$0.12 to  $\Gamma$=2.54$\pm$0.07), a BLLac intermittently exhibiting broad emission lines  characteristic of FSRQs \citep{Ver95}. It would be interesting to investigate  the correlation of this spectral evolution with properties in other bands. 

The distributions in Fig. \ref{fig:index} are remarkably narrow, with rms =0.16-0.18, i. e. comparable to the index statistical uncertainties for the faintest  LBAS sources. Note that the overlap between LSP-BLLac and FSRQ photon index  distributions is small (LSP-BLLacs being harder), although both have similar positions  of their synchrotron peaks.  The gamma-ray photon index is thus a distinctive  property of a blazar subclass.  The low dispersion observed within a subclass  strongly supports the idea that a very limited number of physical  parameters (possibly only one), drive the spectrum shape in the GeV energy range.  It can also be connected to distinct dominant emission mechanisms for the  different classes, External Compton for the low-energy peaked sources and  Synchrotron Self Compton for the high-energy peaked ones, as discussed in  \cite{SEDpaper}.
 
\begin{figure}
\epsscale{0.6}
\plotone{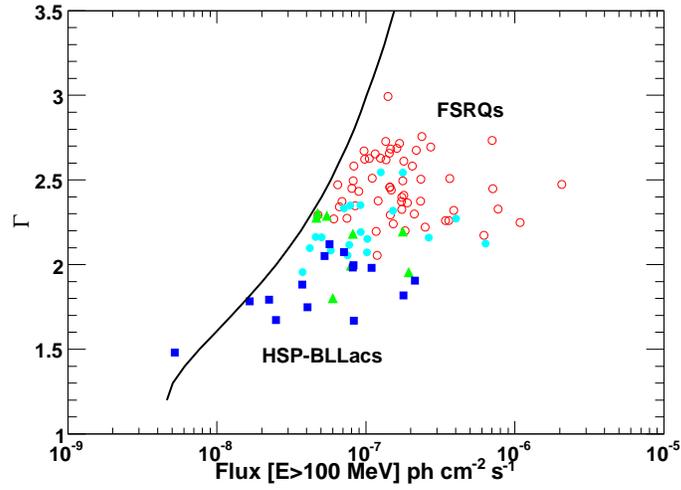}
\caption{Photon index $\Gamma$ vs Flux ($E>100$ MeV) for the LBAS sources 
considered here. Open circles, red: FSRQs; solid symbols: BL Lacs (cyan circles: LSP-BLLacs, green triangles: ISP-BLLacs, blue squares: HSP-BLLacs). The solid curve represents the $\mathrm{TS}=100$ limit for a 3 month period (i.e. the defining condition of the LBAS sample) estimated for (l,b)=($80^\circ$, $40^\circ$).}
\label{fig:flux_index}
\end{figure}

One must keep in mind that the LBAS sample, being significance limited (TS$>$100  after 3 months of LAT operation) has an intrinsic bias such that faint sources  can more easily be detected if they are hard. This situation is illustrated in  Fig. \ref{fig:flux_index} \citep[similar to Fig. 7 in ][]{LBAS}, where the 6-month  average photon index flux is plotted vs the corresponding flux for the four  subclasses along with the approximate LBAS flux limit (for the first 3 months of  operation).  From this figure, the source subclass that appears potentially the  most affected by this bias is that of HSP-BLLacs, which are substantially fainter than  the other sources. However, as pointed out in \cite{SEDpaper}, more than 60\% of  known radio-loud HSP-BLLacs are included in the LBAS sample, so the measured photon  index distribution is probably still representative of the full population.

\subsection{Photon index variability}
 
\begin{figure}
\epsscale{1.}
\plotone{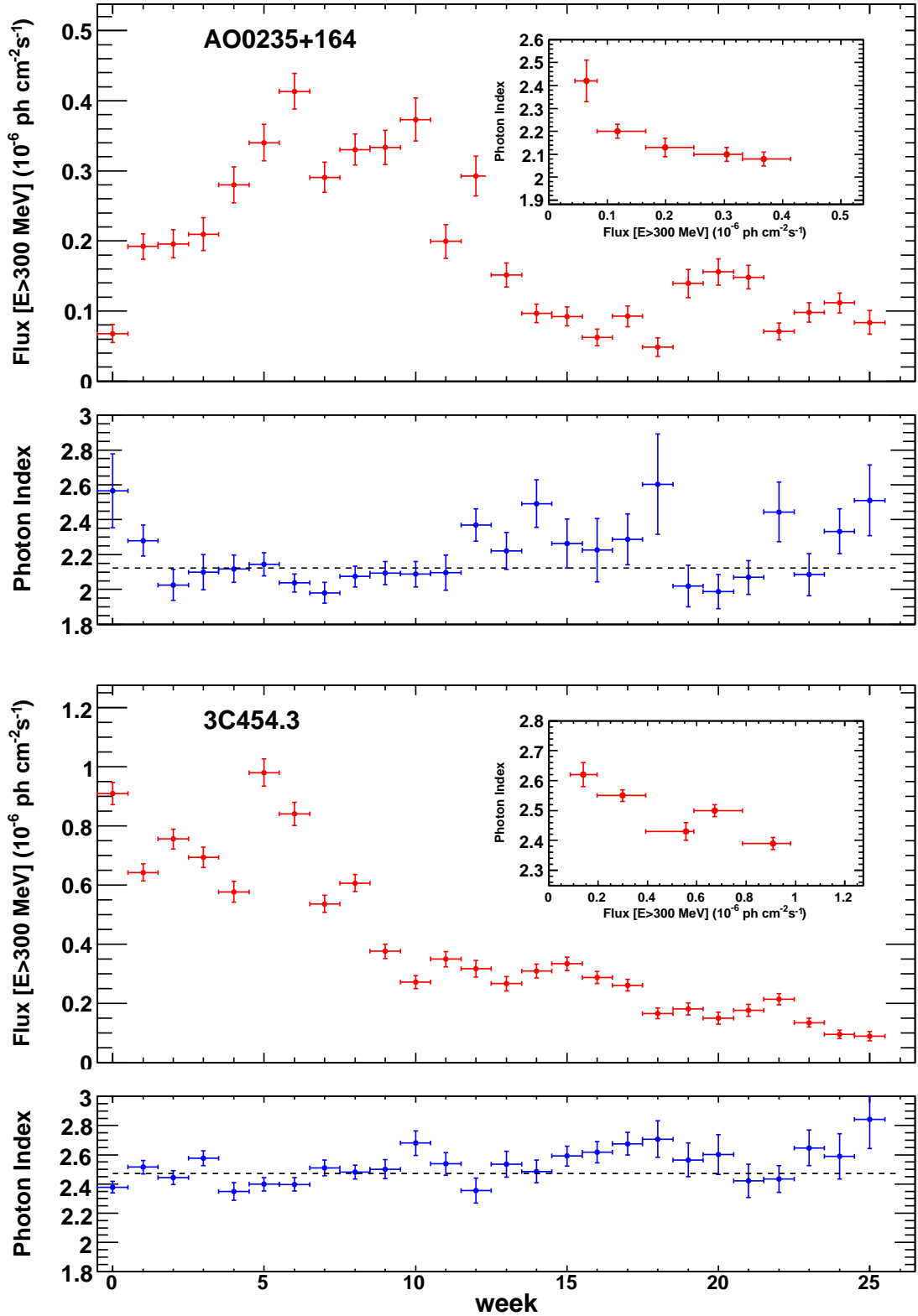}
\caption{Measured weekly fluxes and photon index for 3C\,454.3 and AO\,0235+164 used as input of the simulations described in the text. The insets show the photon index resulting from an analysis where photons were sorted in five weekly-flux bins plotted vs the weekly flux.}
\label{fig:lc_flux_index}
\end{figure}

Given the narrowness of the photon index distribution for a given class, the photon index would not be expected to vary wildly over time for a given source.   Two examples of weekly flux and photon index light curves are shown in Fig.  \ref{fig:lc_flux_index} for the brightest FSRQ (3C\,454.3) and LSP-BLLac (AO\,0235+164) in  the LBAS. The average photon index is shown as a dashed line in the  corresponding panels.  Although flux variations are large (flux variations by a  factor $>$7) for both sources, the range of photon index values is only about  0.3 wide if one allows for a one-sigma dispersion.  The insets in Fig.  \ref{fig:lc_flux_index} display the photon index resulting from an analysis  where photons were sorted in five bins in weekly flux, plotted vs the weekly  flux. A weak ``harder when brighter'' effect can be seen for both sources. 

\begin{figure}
\epsscale{0.8}
\plotone{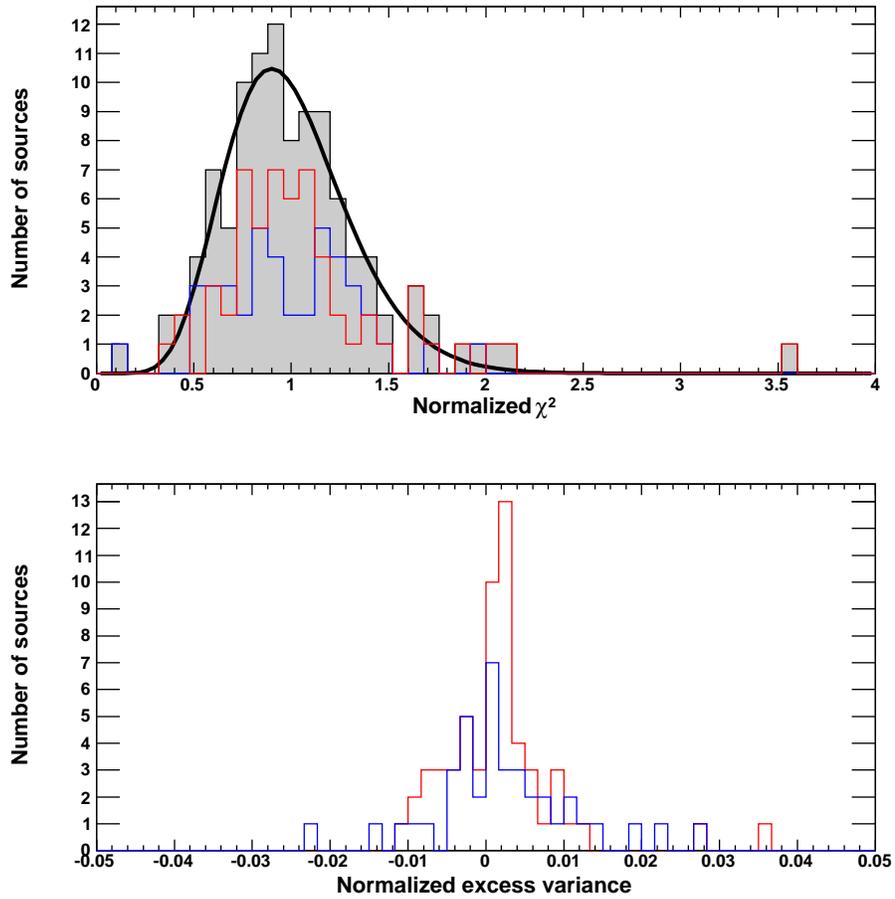}
\caption{Top: Normalized $\chi^2$ distributions of the weekly photon indices for FSRQs (red), BLLacs (blue) and all sources (black). Bottom: Normalized excess variance distributions of the weekly photon indices for FSRQs (red) and BLLacs (blue).} 
\label{fig:index_variability}
\end{figure}

\begin{figure}
\epsscale{0.8}
\plotone{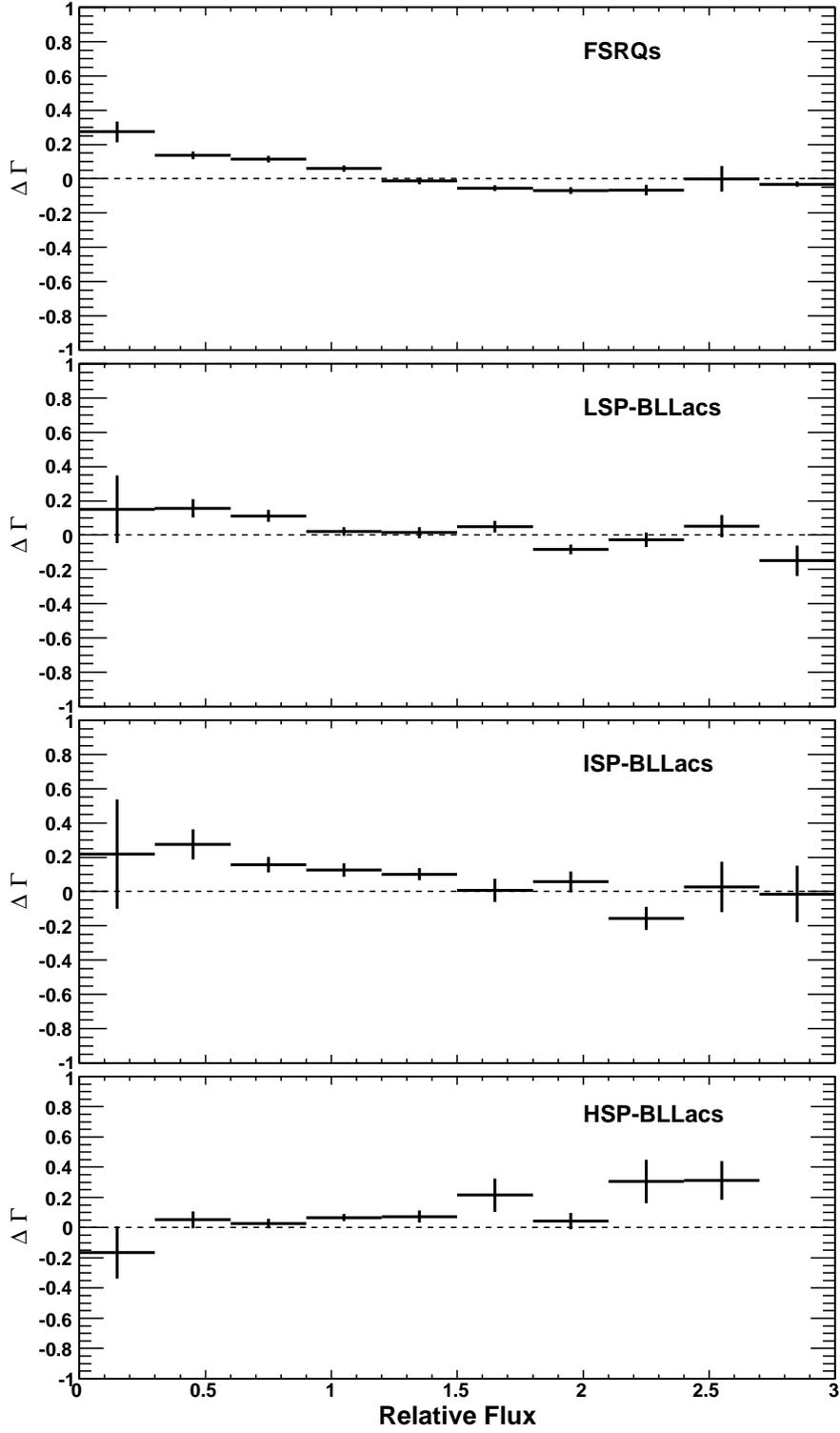}
\caption{Average variation of photon index, $\Gamma$, vs relative flux 
(normalized to the average flux), for the different blazar subclasses. The data 
are integrated over a weekly timescale, and only the eight brightest 
representatives of each subclass have been considered. }
\label{fig:flux_vs_index}
\end{figure}

To test the constancy of the photon index, the weekly photon indices were fitted  with a constant model and the corresponding fractional excess variance  \citep{Vaugh03} was calculated for the different LBAS sources, keeping only time  periods where the sources were detected with a significance larger than 3  $\sigma$. Fig. \ref{fig:index_variability} top displays the distributions of  normalized $\chi^2$ values obtained for FSRQs (red), BLLacs (blue) and all  sources (black). The means of these distributions are close to 1, as expected  for a constant photon index, with no significant differences between FSRQs and  BLLacs.  For illustration, the normalized $\chi^2$ distribution with 20 degrees  of freedom (corresponding to an average source) expected in the case of a  constant photon index is plotted as well.  The source associated with a  normalized $\chi^2$ of 3.5 is PKS\,1502+106 for which a clear indication of  ``harder when brighter'' effect has been observed during a bright flare  \citep{Cip09}. \mbox{Fig. \ref{fig:index_variability}} bottom shows the  distributions of fractional excess variance for FSRQs and BLLacs.  The means of  the distributions are compatible with 0 (within 1 $\sigma$); the same applies  for all three BLLac subclasses.  The widths of these distributions are in good  agreement with the average statistical uncertainties \citep[estimated from Eq. 11 in][]{Vaugh03}, i.e. 8$\times10^{-3}$ and 1.5$\times10^{-2}$ for FSRQs and BLLacs  respectively. Inspection of distributions of Pearson coefficients of the weekly  flux vs photon index correlation does not reveal any strong trend for any  subclass.  The average variation of weekly photon index is plotted vs the  relative flux (normalized to the average flux) in Fig. \ref{fig:flux_vs_index}  for the different blazar subclasses.  The data from the eight brightest  representatives of each subclass have been considered\footnote{A condition on  the significance $> 1\sigma$, i.e. less restrictive than above, has been imposed  in this analysis to minimize the effect of the instrumental bias against soft  and faint states illustrated in Fig. \ref{fig:flux_index}.}.  A weak ``harder  when brighter'' effect is apparent in FSRQs, LSP-BLLacs and ISP-BLLacs, whereas no  significant effect is present for HSP-BLLacs\footnote{The weak tendency toward softer spectra at high flux in HSP-BLLacs does not appear to be significant.}.

These observations, aimed at determining the gross spectral features of a source  ensemble, do not exclude fine spectral evolutions over short periods of time,  e.g. regarding particular episodes of flaring activity, or for particular  sources.  They do, however, demonstrate that the photon index in the GeV range changes  little with time and within a blazar subclass.

\subsection{Spectra of brightest sources} 

\begin{figure}
\epsscale{1.}
\plotone{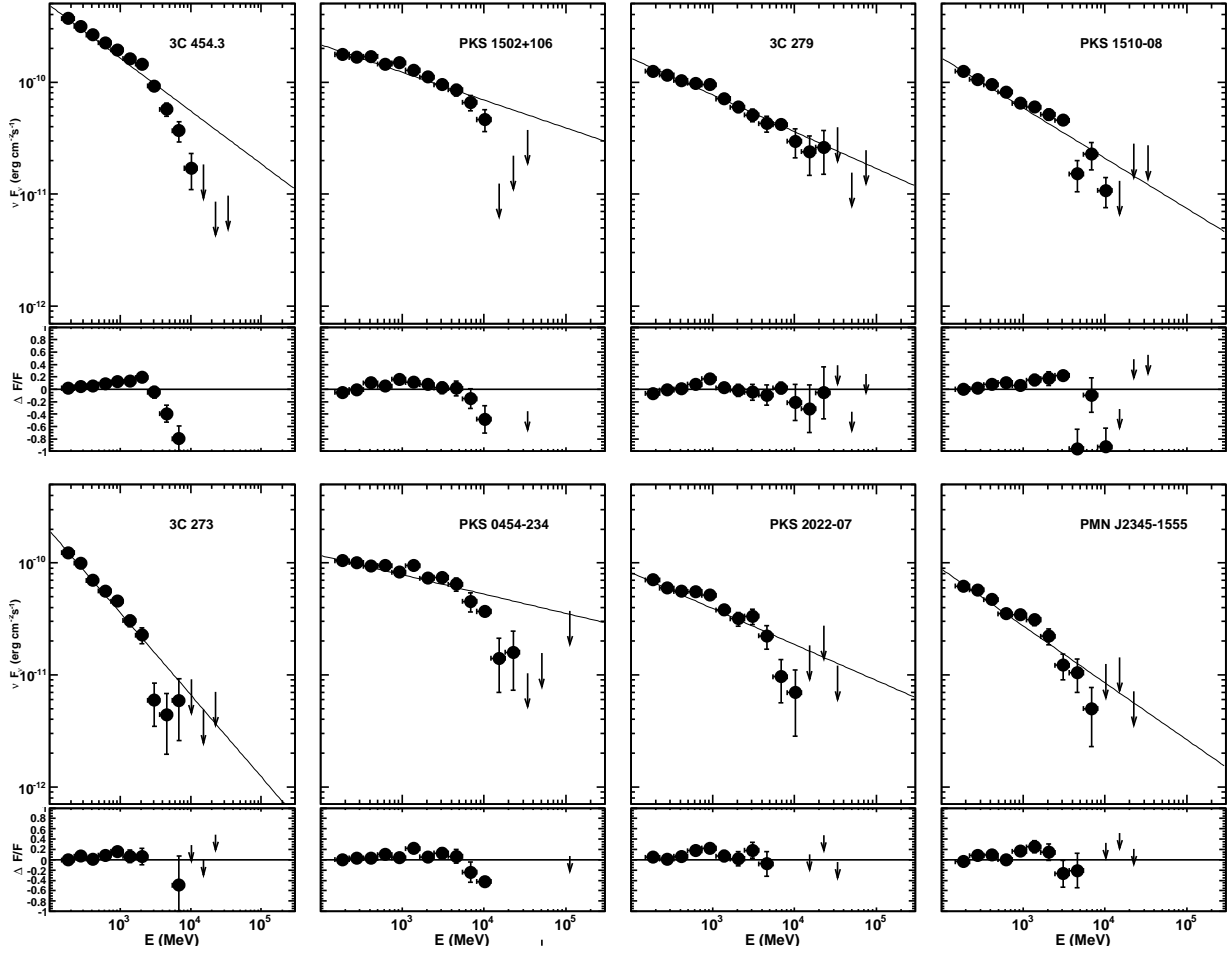}
\caption{Gamma-ray spectra of the eight brightest FSRQs in the LBAS sample obtained for equispaced logarithmic bins (dots), together with residues with respect to the fitted power-law model (solid lines in upper panels).}
\label{fig:sed_fsrq}
\end{figure}

\begin{figure}
\epsscale{1.}
\plotone{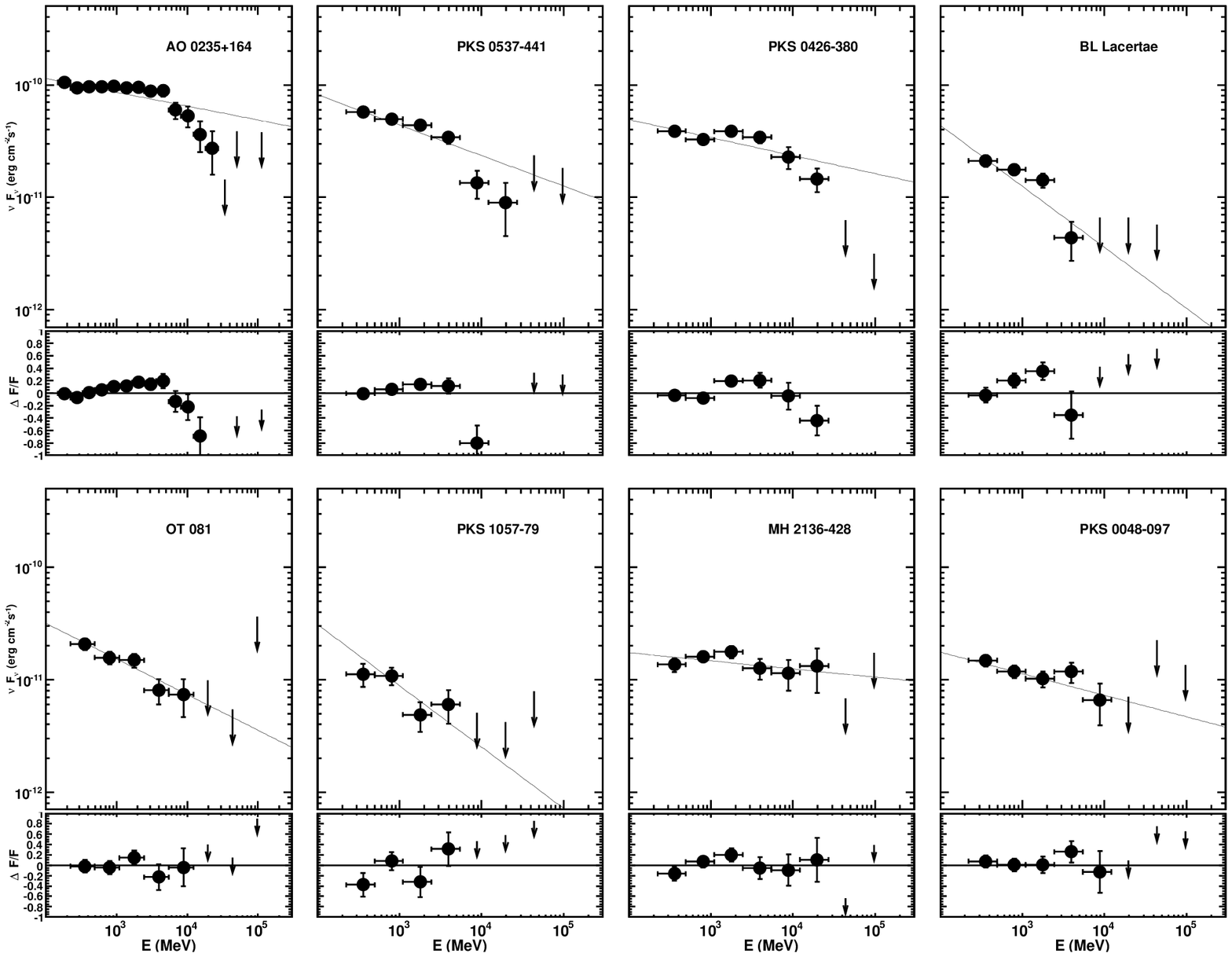}
\caption{Same as Fig. \ref{fig:sed_fsrq}, for the eight brightest LSP-BLLacs in the LBAS sample.}
\label{fig:sed_lbl}
\end{figure}

\begin{figure}
\epsscale{1.}
\plotone{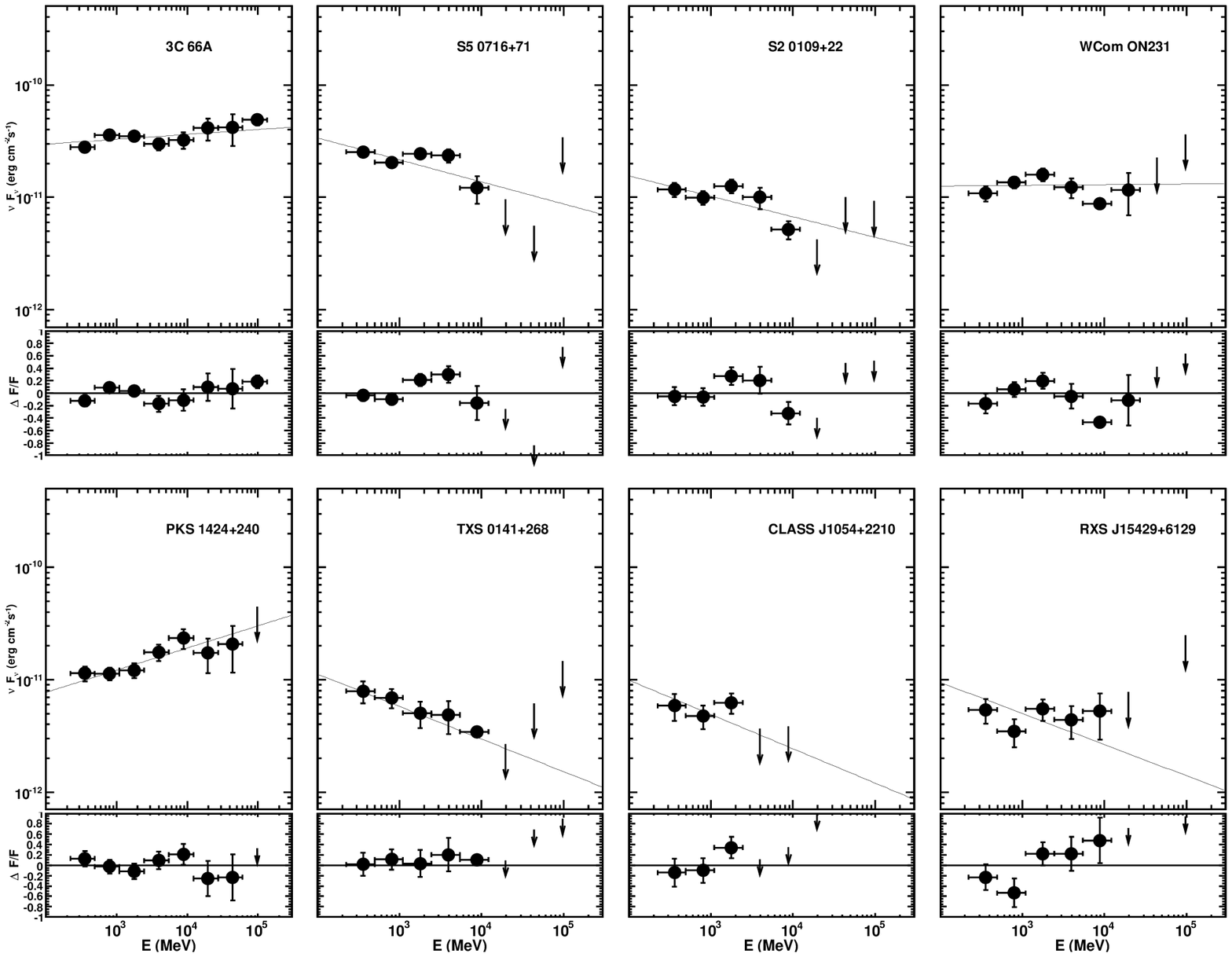}
\caption{Same as Fig. \ref{fig:sed_fsrq}, for the eight brightest ISP-BLLacs in the LBAS sample.}
\label{fig:sed_ibl}
\end{figure}

\begin{figure}
\epsscale{1.}
\plotone{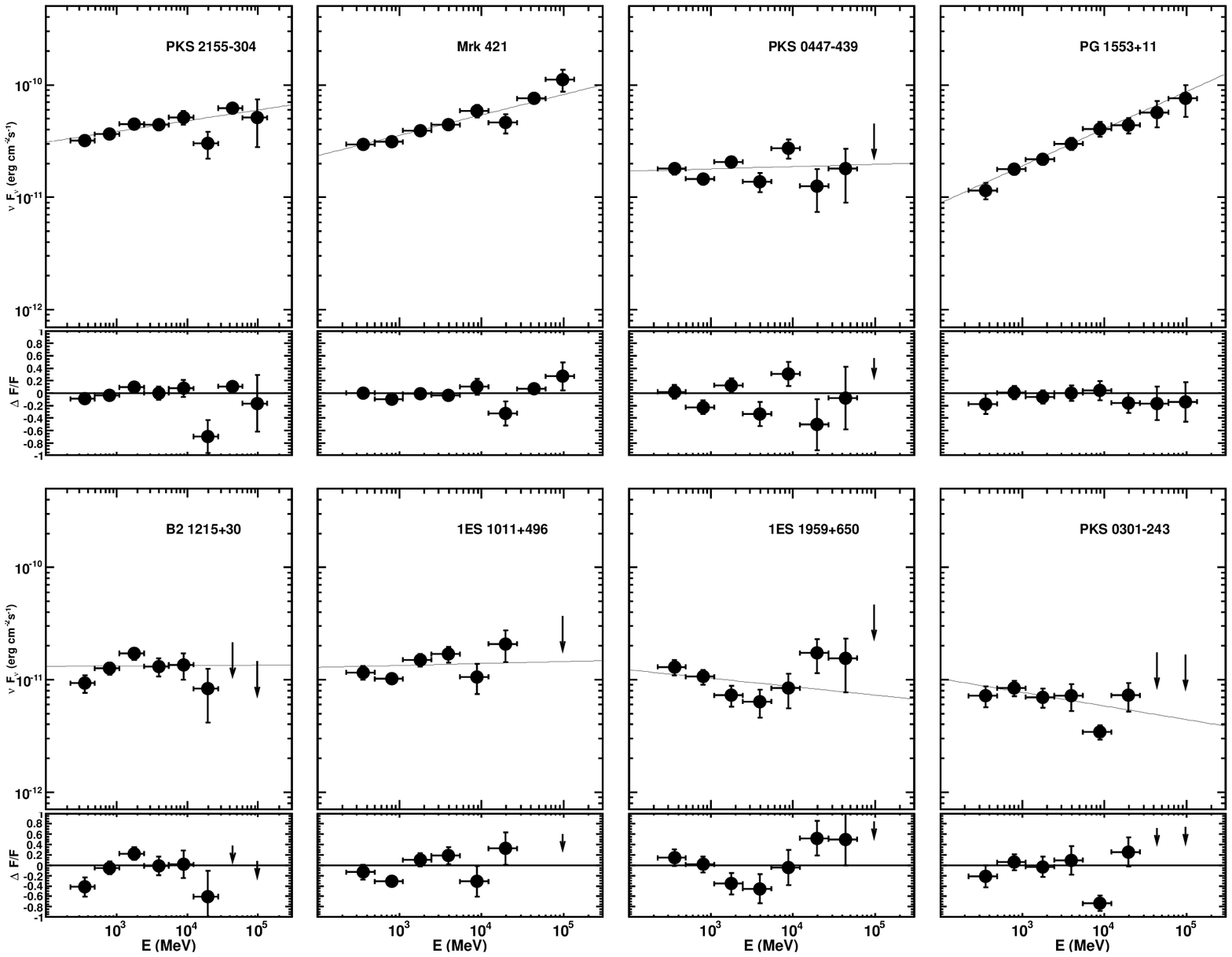}
\caption{Same as Fig. \ref{fig:sed_fsrq}, for the eight brightest HSP-BLLacs in the LBAS sample.}
\label{fig:sed_hbl}
\end{figure}

\begin{figure}
\epsscale{1.}
\plotone{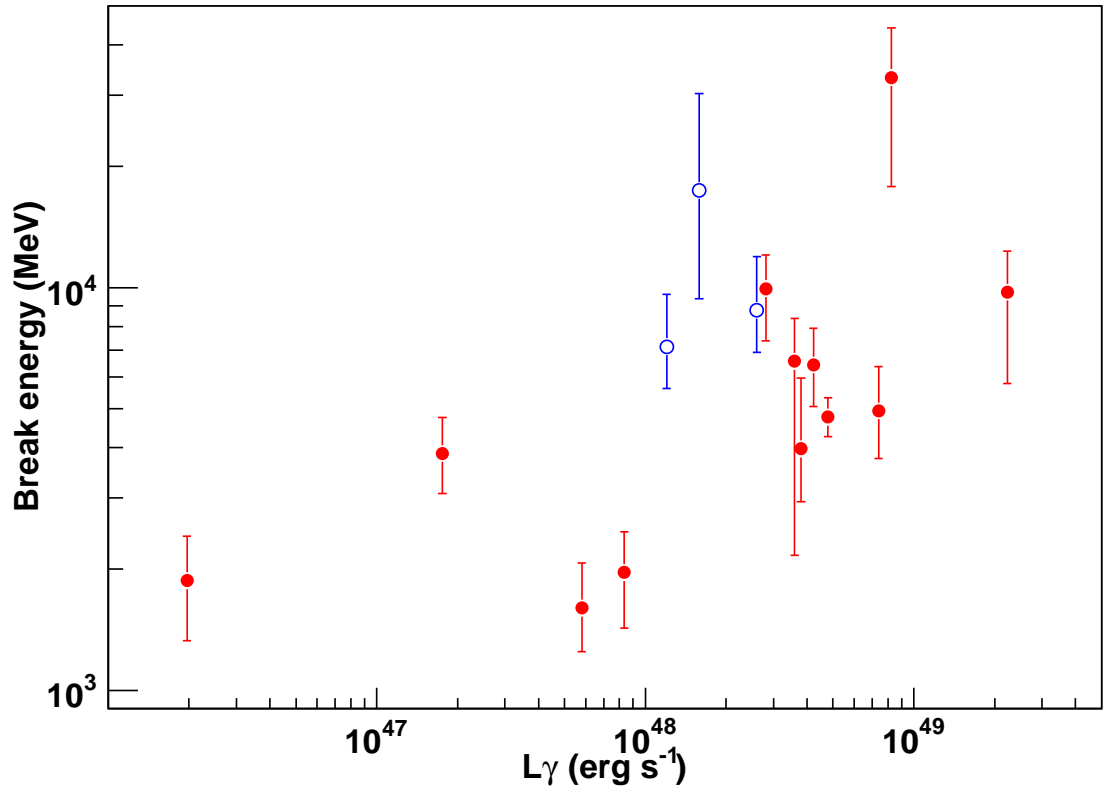}
\caption{Break energy vs gamma-ray luminosity for the FSRQs (closed red symbols) and LSP-BLLacs (open blue symbols) listed in Table 1 and 2 respectively.}
\label{fig:break_lum}
\end{figure}

\begin{figure}
\epsscale{1.}
\plotone{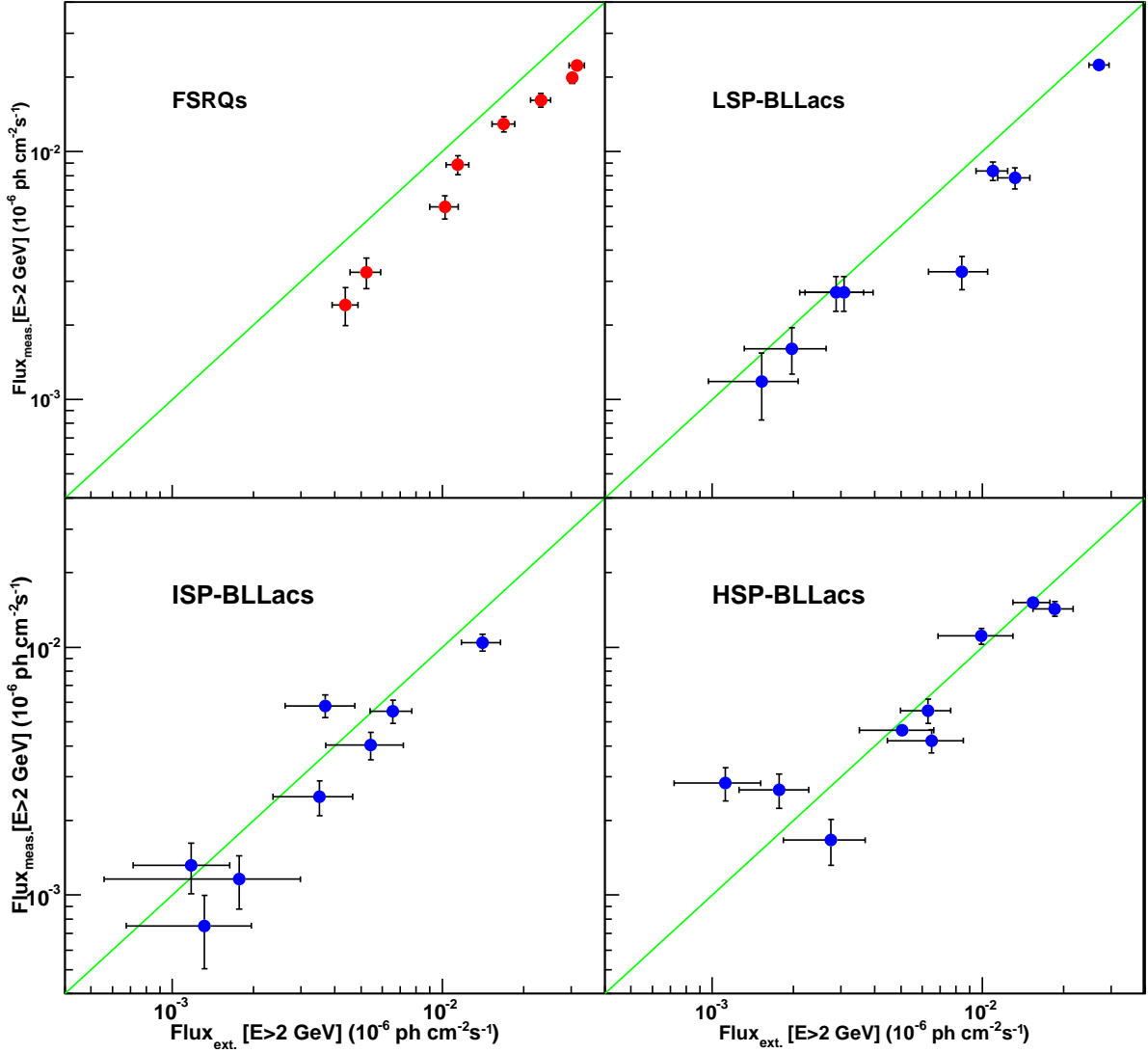}
\caption{Measured vs extrapolated flux above 2 GeV for the eight brightest sources of each blazar subclass. The distance of the points from the diagonal is indicative of the presence of a spectral break.}
\label{fig:two_bands}
\end{figure}

In this section we present the spectra obtained for the eight brightest  representatives of the four subclasses, FSRQs (Fig. \ref{fig:sed_fsrq}), LSP-BLLacs  (Fig. \ref{fig:sed_lbl}), ISP-BLLacs (Fig. \ref{fig:sed_ibl}), and HSP-BLLacs (Fig.  \ref{fig:sed_hbl}), ordered according to decreasing average flux.  Upper limits  are shown for bins associated with a TS lower than 9 (significance lower than 3  $\sigma$) or with a number of source photons (predicted by the model) lower than  3.  The brightest FSRQ, 3C\,454.3 with an average flux (E$>$100 MeV) of 2$\times  10^{-6}$ ph cm$^{-2}$s$^{-1}$, exhibits a pronounced break around 2 GeV, as  reported in \cite{LAT3C454.3}. Indications for breaks between 1 and 10 GeV are  observed for essentially all of these FSRQ sources.  This behavior is confirmed  by comparing (Fig. \ref{fig:two_bands}) the flux (E$>$2 GeV) extrapolated from  the spectral distribution in the range 100 MeV $<$E$<$ 2 GeV, fitted with a  power-law function, with the actual measured flux (obtained via a power-law fit  in the E$>$2 GeV energy range).  For all sources considered, the measured flux  is lower than the extrapolated flux by more than 30\%.  The spectral properties  of these sources are summarized in Table 1, which gives also the difference in  loglikelihood between the BPL and PL fits, the break energy in the source rest  frame, E$^\prime$$_{Break}$=E$_{Break}\times$(1+z), and the source gamma-ray  luminosity \citep[computed as in][]{Ghi09}.  E$^\prime$$_{Break}$ lies between 1.6  and 10 GeV for all sources listed in Table 1. No significant correlation is  found between E$^\prime$$_{Break}$ and the gamma-ray luminosity (Fig. \ref{fig:break_lum}).

The presence of a break is also clear for two of the brightest LSP-BLLacs (associated  with the highest luminosities, Table 2) in the LBAS sample, AO\,0235+164 and  PKS\,0537-441 (Fig. \ref{fig:sed_lbl}), while it is less apparent for the  fainter ones.  From the inspection of these spectra, it can already be noted  that the onset of the break, when present, seems to be located at higher energy  than in the case of FSRQs while E$^\prime$$_{Break}$ lies in about the same  range.

Some ISP-BLLacs (Fig. \ref{fig:sed_ibl}) present clear signs of breaks (e.g.  S5\,0716+71, S2\,0109+22), while the rest, having in some cases very hard  spectra, are compatible with power-law distributions up to several tens of GeV.   Finally, no bright HSP-BLLac (Fig. \ref{fig:sed_hbl}) shows any evidence for a break  in the LAT energy range.  This simple observation, which could be naively  expected for sources many of which are detected at TeV energies, definitely  rules out an instrumental effect as the origin of the break found for lower-energy peaked sources. The indication for a moderate break in PKS\,2155-304  \citep{Aha09} observed over a short time period (11 days) does not persist over  a 6-month integration time. 

 For a small number of sources exhibiting the break, a few photons compatible  with the source location are detected at high energy, at variance with the  decreasing trend.  Most of these have been cut off by the condition on the minimum number of photons per bin (3).  More statistics will be required  to determine whether these photons do arise from these sources or just represent background fluctuations. 

Fig. \ref{fig:two_bands} illustrates the general trend for the four subclasses.  The trend observed in the presence of a spectral break (or curvature) parallels  that observed in the photon index for the four different classes.

\subsection{Spectra of special sources}
 
\begin{figure}
\epsscale{1.}
\plotone{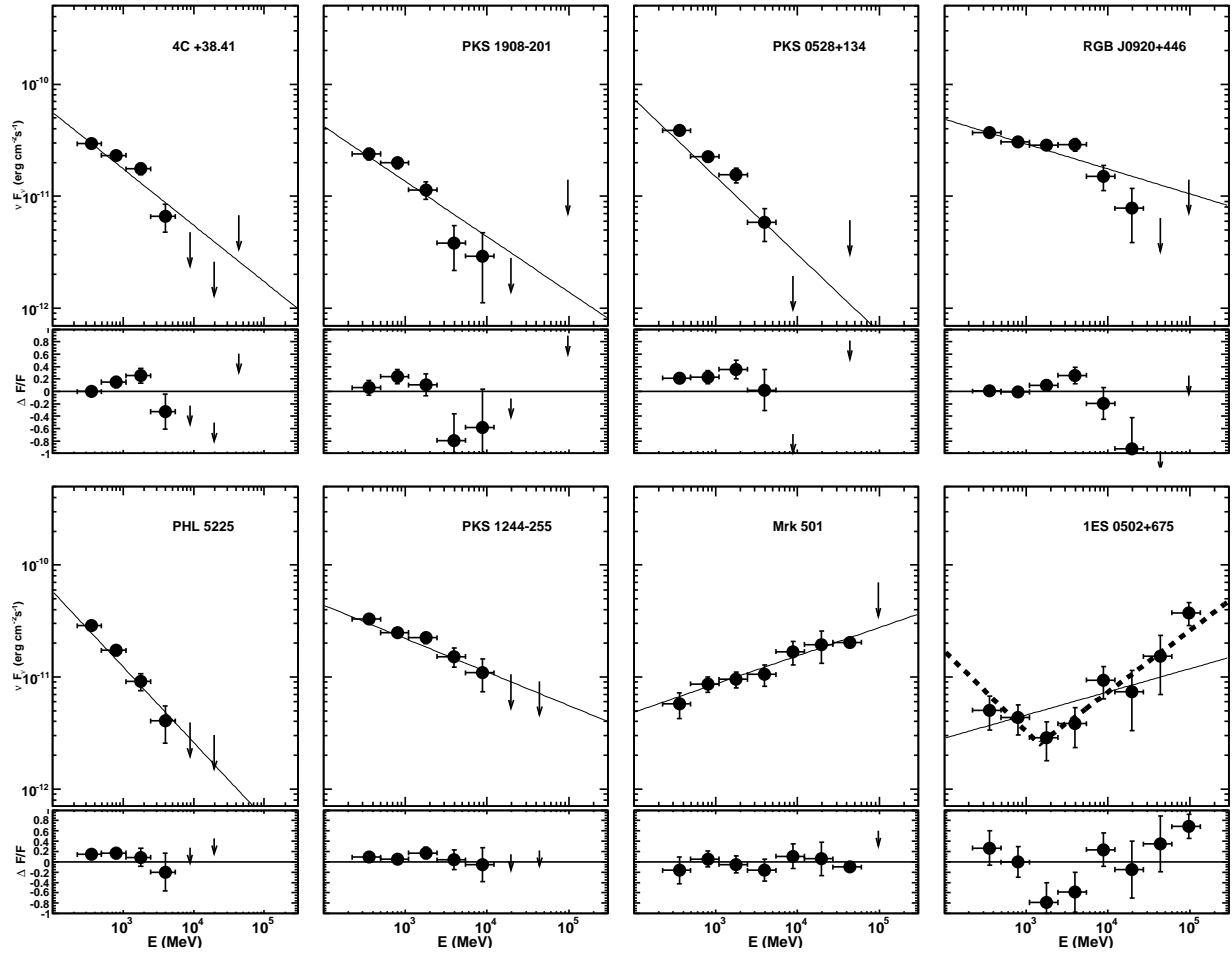}
\caption{Same as Fig. \ref{fig:sed_fsrq}, for eight particular sources in the LBAS sample.}
\label{fig:sed_spec}
\end{figure}

For 19 of the 22 brightest LBAS FSRQs, a likelihood ratio test \citep[LRT,][]{mat96} rejects the  hypothesis that the spectrum is a PL (null hypothesis) against the one that the  spectrum is a BPL, at a confidence level greater than 97\%.  The four top panels  of Fig. \ref{fig:sed_spec} show spectra of representative sources where the  break is clear. The spectral properties of these sources (PL and BPL fit  results, difference in logLikelihood between the two fits) are reported in Table  1 as well.  The two bottom left panels correspond to two of the three sources  having the confidence level less than 97\% (the other being PKS\,2022-07, whose  spectrum is given in Fig. \ref{fig:sed_fsrq}).  For these two sources, a break  located around 10 GeV cannot be excluded.

The last two panels of Fig. \ref{fig:sed_spec} present the energy spectra of  sources of particular interest, Mrk\,501 and 1ES\,0502+675. Mrk\,501, the archetypal  example of an extreme HSP-BLLac, is hard ($\Gamma$=1.75$\pm$0.06) and does not exhibit  any sign of curvature, in keeping with the behavior of the other HSP-BLLacs. The other  one corresponds to 1ES\,0502+675, an HSP-BLLac which exhibits an unusual concave energy  spectrum ($\Gamma_1$=2.68$\pm$0.18,$\Gamma_2$ =1.47$\pm$0.10, E$_{break}$=  1.4$\pm$0.6 GeV).  The LRT indicates that a BPL  model is favored against a PL with a significance level of $2\times10^{-4}$.  The very hard spectrum above 1.4 GeV would make  this source a prime target for TeV observations although the redshift is fairly  large (0.416).
 
\subsection{Detailed analysis of 3C\,454.3 and AO\,0235+164 energy spectra}

\begin{figure}
\epsscale{1.}
\plotone{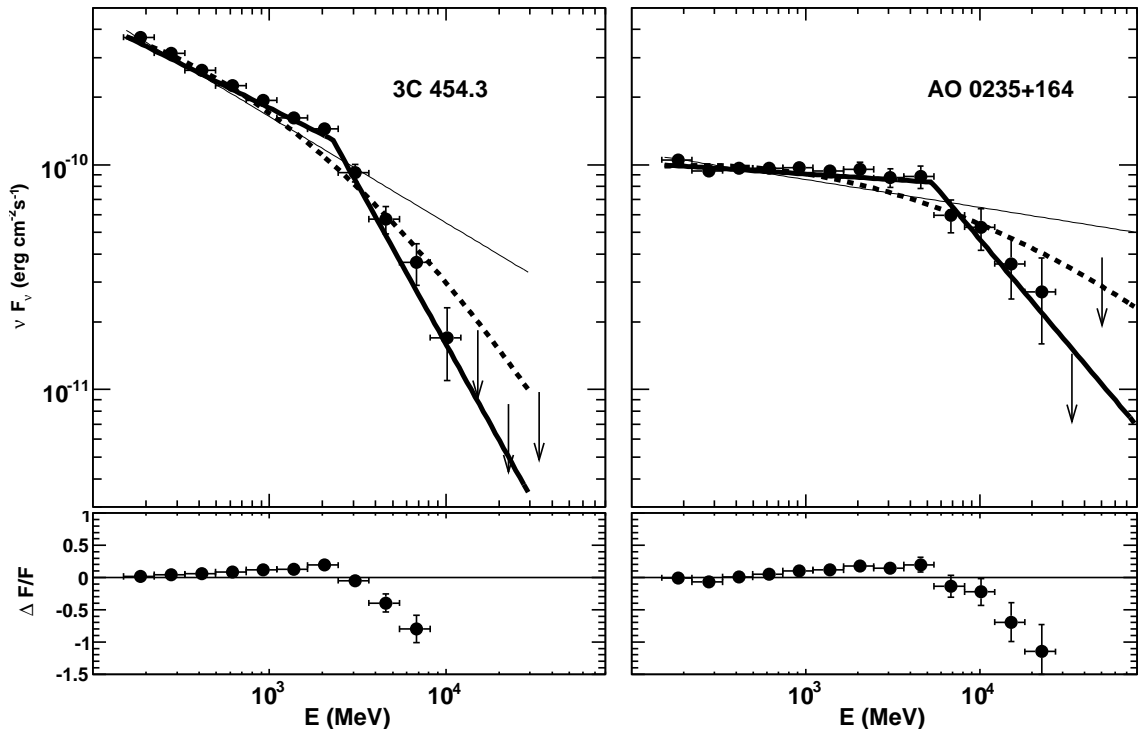}
\caption{Upper panels: Energy spectra of 3C\,454.3 (left) and AO\,0235+164 (right) compared with fit results obtained with different models: PL (thin solid), BPL (thick solid) and logparabola (dashed).}
\label{fig:zoom}
\end{figure}

The data accumulated over 6 months enable us to discriminate between different  spectral models for the two brightest sources with spectra exhibiting strong  departure from a pure PL, namely 3C\,454.3 and AO\,0235+164.  Fig.  \ref{fig:zoom} shows the results of fits with different models: PL (thin lines),  BPL(thick solid) and log-parabola (dashed) are compared with the data.   Surprisingly, a broken power-law model is favored as the best fit for both  sources.  Despite the good statistics, no curvature is apparent in the energy  range below the break.  The fitted parameters are:  
F[E$>$100 MeV]=1.97$\pm$0.03$\times 10^{-6}$ ph cm$^{-2}$s$^{-1}$, 
$\Gamma_1$=2.39$\pm$0.02, $\Gamma_2$=3.42$\pm$0.11, E$_{break}$= 2.5$\pm$0.3 GeV  for 3C\,454.3 and F[E$>$100 MeV]=0.60$\pm$0.02$\times 10^{-6}$ ph cm$^{-2}$s$^{-1}$,  $\Gamma_1$=2.05$\pm$0.02, $\Gamma_2$=2.95$\pm$0.16, E$_{break}$=4.5$^{+1.5}_{-1.0}$GeV for AO\,0235+164. While the break energy is somewhat larger for  AO\,0235+164 than for 3C\,454.3, the photon index change ($\Gamma_2$ - $\Gamma_1$)  is about the same (0.90$\pm$0.16 vs 1.03$\pm$0.11). The similarity of the break feature for two sources belonging to  different subclasses, FSRQ and LSP-BLLac, with different line strengths, seems to  rule out any absorption effect in the LSP-BLLacs.

\subsection{Apparent curvature due to varying spectral hardness}

\begin{figure}
\epsscale{1.}
\plotone{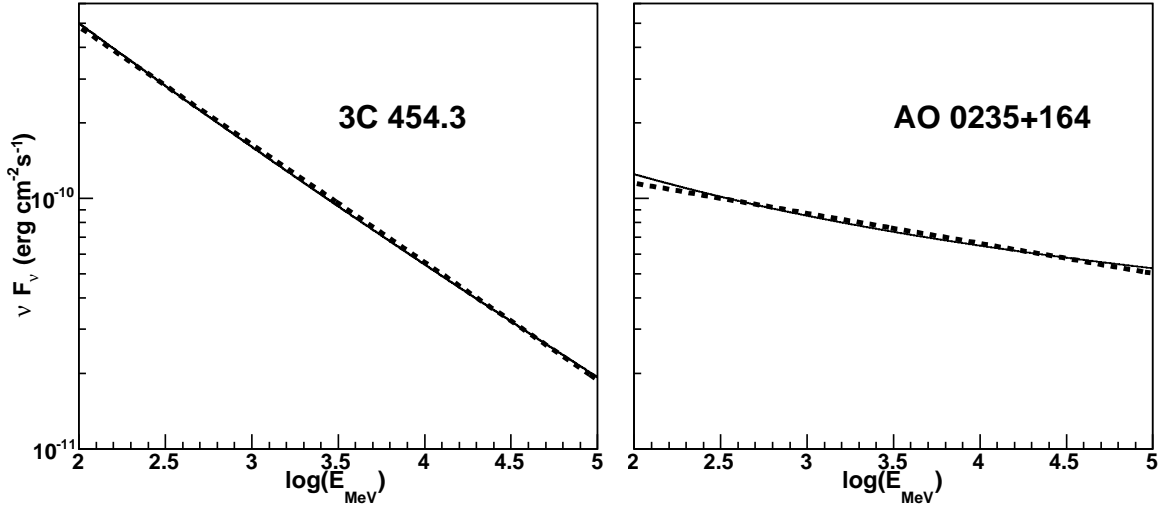}
\caption{Analytical energy spectra (solid curves) of 3C\,454.3 and AO\,0235+164 resulting from summing power-law distributions with parameters (flux, photon index) as measured in weekly bins (Fig. \ref{fig:index_variability}). The dashed lines represent PL fits.}
\label{fig:ana}
\end{figure}

\begin{figure}
\epsscale{1.}
\plotone{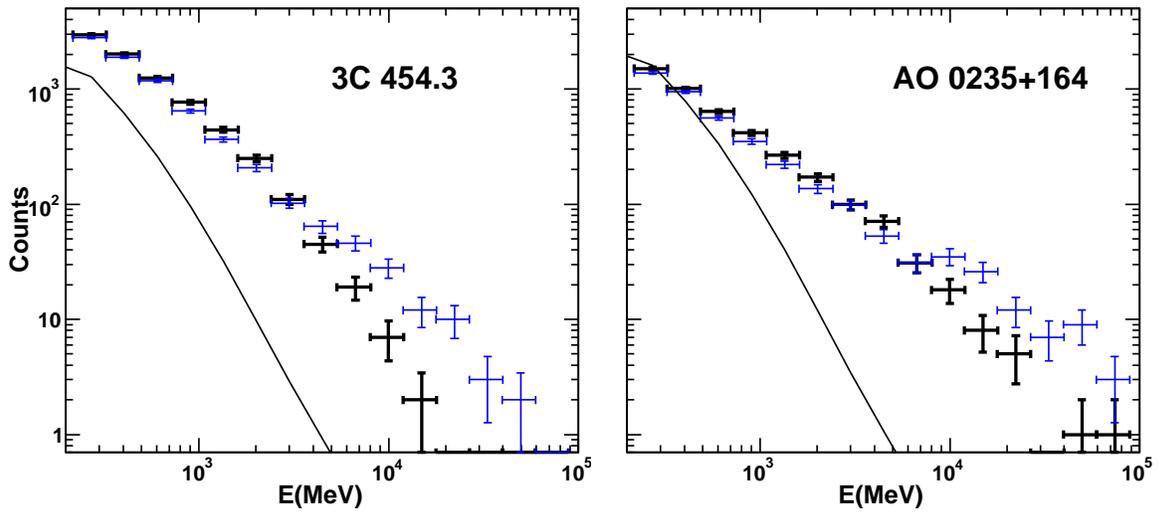}
\caption{Count distributions within the 90\% containment radius of simulated (blue, thin) and real (black, thick) data for 3C\,454.3 and AO\,0235+164. The solid curve correspond to the total contribution of galactic and isotropic diffuse backgrounds. }
\label{fig:res_data_sim}
\end{figure}

An energy spectrum with time varying hardness may exhibit an apparent curvature  when integrated over an extended period of time. To assess the magnitude of this  effect, both analytical estimates and simulations assuming pure power-law  distributions with flux and photon index corresponding to those actually  measured over weekly time bins have been performed. Fig. \ref{fig:lc_flux_index}  shows the 3C\,454.3 and AO\,0235+164 weekly light curves and corresponding  photon index.  With these input data, the calculated spectra (assuming constant  exposure) are not found to exhibit significant curvatures (Fig. \ref{fig:ana}).   Fig. \ref{fig:res_data_sim} compares the simulated photon count distribution  obtained within the 90\% containment radius around the source (blue) with the  data (black). The effect of spectral hardness varying with time clearly cannot  alone account for the observed features.

\section{Discussion}

The trend in the observed gamma-ray photon index reported here confirms that  reported earlier using three months of data: FSRQs, with a gamma-ray photon  index greater than 2, are softer than BLLacs, indicating that the peak of the  high-energy component in FSRQS is always lower than 100 MeV.  For BLLacs, the  gamma-ray photon index correlates with the different BLLac subclasses, which  themselves are defined by the position of the synchrotron peak.  The measured  photon index shifts from $\Gamma >$2 to $\Gamma <$2, indicating that the peak  energy of the high-energy component of the spectral energy distribution (SED)  sweeps across the Fermi energy range from LSP-BLLacs to HSP-BLLacs.

The photon index being related to the shape of the emitting electron energy  distributions, different regions of the electron distributions (from the high- to the low-energy ends for FSRQs/LSP-BLLacs and HSP-BLLacs respectively) are probed in the  LAT range. One would thus expect different spectral variability patterns in the  LAT energy range for different blazar classes.
 
Using weekly light curves obtained over the first 6 months of LAT operation,  this expectation does not seem to be corroborated by the data. The gamma-ray  photon index appears remarkably stable with time, irrespective of the blazar  class. This feature is in line with the observed narrowness of the index  distribution for a class: a strongly varying photon index for a source would  inevitably lead to a broad class distribution as different sources would be  ``caught'' in different states. This apparent constancy of the photon index for  all blazar classes may appear surprising, since for different classes, different  electron energies (potentially associated with different cooling timescales,  $t_{\rm cool}$) emit gamma-rays in the LAT energy range. However, estimates of the cooling time show that within one week, all pairs with  Lorentz factors $\gamma > 5/\delta_{10} u'$ have cooled down (where  $\delta_{10}$ is the Doppler factor $\delta/(1+z)$ divided by 10, and $u'$ is  the jet frame energy density in erg cm$^{-3}$ of the ambient magnetic field, or  photon field provided scattering occurs in the Thomson regime).  For  sufficiently large $u'$ the LAT energy range falls in the complete cooling  regime.  Spectral hysteresis may then be expected on time scales shorter than  weekly if the duration of electron injection is sufficiently limited $\ll t_{\rm  cool}$. Flux variations on weekly time scales may then reflect a varying  injected energy content into the emission region from one week to another.   Alternatively, continuous particle injection on at least weeks time scale could  stabilize the spectral index. In the case of lower $u'$, constant spectral  indices can still be expected during flux variations in the LAT energy range if  continuous particle injection is maintained for a time range significantly  longer than the cooling time of the gamma-ray emitting particles \citep[e.g.][]{Kir99}. Flux changes then constrain the duration of particle injection.  As more data are accumulated, possible exceptions to the observed index  stability may appear. However, we can safely claim that a very soft spectrum for  an HSP-BLLac, like that reported from EGRET for PKS\,2155-304, $\Gamma$=2.35  \citep{3EGcatalog}, or a very hard one for a FSRQ, like the one reported from  AGILE for 3C\,454.3, $\Gamma$=1.7 \citep{Ver09}, represent very rare  occurrences. 

All blazar spectra measured by EGRET were represented with pure PL. Thanks to  its improved sensitivity, Fermi has revealed that the spectra of some low-energy  peaked blazars display strong departure from a pure-PL behavior, with a BPL  function as the best model. This feature being present for essentially all FSRQs  where it can be detected with sufficient significance, it is likely to be a  general character of this class. The three brightest LSP-BLLacs, in which this effect  is also clearly seen, have higher luminosity (L$_{\gamma}>$10$^{47}$ erg s$^{-1}$) than the rest of the LBAS LSP-BLLacs. Two of them are known to exhibit broad  emission lines in low emission states, so could be FSRQs whose lines are hidden  by non-thermal emission in active states. As discussed in \cite{LAT3C454.3}, the  difference in photon index for most sources is significantly larger than 0.5  expected for an incomplete cooling effect.  An absorption effect seems to be  ruled out as well, since to produce a break in the 1-10 GeV, the photon field  should have an energy peaking in the 0.05-0.5 KeV range, which excludes the  broad-line region peaking in the UV. This feature most likely reflects the  energy distribution of the emitting electrons.  For the low-energy peaked  sources where it is seen, $\Gamma$ is greater than 2, i.e. the LAT range  corresponds to``the falling edge'' of the IC hump, where the highest energy  electrons contribute. This feature could indicate a cutoff in that distribution,  possibly related to limitations in the acceleration process \citep[e.g.][]{Dru91, Web84}. Comparison of the maximum electron energies determined by  the SED for PKS\,2155-304, an HSP-BLLac where a one-zone synchrotron/SSC model was  applied to derive the magnetic field and Doppler factor, showed that such an  interpretation requires that the acceleration rate be approximately 3 orders of  magnitude smaller than the maximum acceleration rate determined by the Larmor  timescale \citep{Fin08}. A Klein-Nishina break, expected in the context of a  dominating External Compton emission process where the electrons upscatter BLR  photons, is predicted to set in around $15 \delta \over \Gamma (1+z)$ GeV, where  $\Gamma$ is the blob bulk Lorentz factor \citep{GT09}, i. e. significantly  higher than the energies found here.

Irrespective of its origin, this feature, common for FSRQs and some LSP-BLLacs, has  important practical consequences. First, it surely complicates the assessment of  EBL attenuation effects using FSRQs and LSP-BLLacs, as fewer photons are detected in  the $>$10 GeV energy range. Second, as the low-energy peaked blazars are likely  to represent the bulk of the blazar population, this break must manifest itself  in (and be considered when evaluating) the contribution of unresolved blazars to  the extragalactic diffuse $\gamma$-ray background. Finally, this effect must be  considered when estimating the detectability of a source in the TeV range.  

The concave shaped SED measured for the X-ray selected object 1ES\,0502+675  (z=0.416) opens interesting questions.  It could potentially be a spurious  feature resulting from the spatial confusion of a hard source with a soft one.  No evidence for a second source has been found using the {\sl gttsmap} tool of  the standard ScienceTools package. The closest CRATES source is 1.5$^\circ$  away. If confirmed, this peculiar spectral shape indicates either two components  (e.g. Synchrotron-Self Compton and External Compton in the context of leptonic  models) in the high-energy SED (however not expected in HSP-BLLacs in the framework of  one-zone leptonic models), or the turnover from the synchrotron to the high-energy component, although the contemporaneous spectral energy distributions  \citep{SEDpaper} obtained during that period does not seem to support that  interpretation. This object should be a prime target for TeV instruments,  opening interesting perspectives for studies on EBL absorption at TeV energies  given its redshift. 

A similar behavior, although less significant, has also been found for  1ES\,1959+650 (Fig. \ref{fig:sed_ibl}) and PG\,1246+586, which is not included  in the LBAS sample but is present in the LAT Bright Source List\footnote{The low  confidence association obtained with 3 months worth of data has by now turned  into a high-confidence one for this source.} \citep{BSL}.  The spectra of other  sources (e.g. W Comae) exhibit a wavy shape possibly indicative of multiple  components as mentioned above. More detailed analysis will be necessary to  resolve this issue.

\section{Conclusion}

The spectral properties of the LBAS blazars in the gamma-ray band, as determined  over the first 6 months of LAT operation, have been presented.  The average  photon index of LBAS blazars are found to be $\Gamma$=2.46 for FSRQs, $\Gamma$=  2.21 for LSP-BLLacs, $\Gamma$=2.13 for ISP-BLLacs and $\Gamma$=1.86 for HSP-BLLacs, with an rms of  0.16-0.18. Spectral breaks have been observed to be common features in FSRQs  (the break energy ranging from 1 to 10 GeV in the source frame for the brightest  sources), and present also in some bright LSP-BLLacs.
The different spectral features reported here represent challenges for  theoretical models aiming at describing the blazar phenomenon. Although the  fairly strong correlation between photon index and blazar class fits well within  pictures \citep{Ghi98, Boe02} where the cooling due to strong ambient radiation  fields (manifesting themselves via the presence of emission lines) limits the  acceleration of particles at high energy, the near constancy of this photon  index with time and flux variation provides new constraints on the emitting  particle dynamics.  Moreover, the fact that spectra for most FSRQs and some LSP-BLLacs  are best modeled by a BPL with a break in the 1-10 GeV range is quite  unexpected, the break representing a distinctive feature of these sources. The  LAT has already revealed novel aspects of gamma-ray blazars and will help refine  the new picture that progressively emerges as more data (both in the gamma-ray  and other bands) are accumulated. 

\section{Acknowledgments}
\acknowledgments The \textit{Fermi} LAT Collaboration acknowledges generous 
ongoing support from a number of agencies and institutes that have supported  both the development and the operation of the LAT as well as scientific data  analysis. These include the National Aeronautics and Space Administration and  the Department of Energy in the United States, the Commissariat \`a l'Energie  Atomique and the Centre National de la Recherche Scientifique / Institut  National de Physique Nucl\'eaire et de Physique des Particules in France, the  Agenzia Spaziale Italiana and the Istituto Nazionale di Fisica Nucleare in  Italy, the Ministry of Education, Culture, Sports, Science and Technology  (MEXT), High Energy Accelerator Research Organization (KEK) and Japan Aerospace  Exploration Agency (JAXA) in Japan, and the K.~A.~Wallenberg Foundation, the  Swedish Research Council and the Swedish National Space Board in Sweden.

Additional support for science analysis during the operations phase is  gratefully acknowledged from the Istituto Nazionale di Astrofisica in Italy and  the Centre National d'\'Etudes Spatiales in France. 
{\it Facilities:} \facility{{\it Fermi} LAT}.


\bibliography{LBAS_spectra.bib}  

\clearpage
\begin{deluxetable}{cccccccccccccc}
\tablecolumns{12}
\tablewidth{0pc}
\tabletypesize{\tiny}
\tablecaption{Spectral properties of selected FSRQs}
\tablenum{1}
\rotate
\tablehead{
\colhead{Name} & \colhead{l} & \colhead{b} & 
\colhead{Flux$_{PL}$\tablenotemark{a}} & \colhead{$\Gamma$} & \colhead{$\Delta 
L$} & \colhead{Flux$_{BPL}$\tablenotemark{a} } & \colhead{$\Gamma_1$} & 
\colhead{$\Gamma_2$} & \colhead{$\Delta \Gamma$} & \colhead{E$_{Break}$ (GeV)} & 
\colhead{z} & \colhead{E$\prime_{Break}$ (GeV)} & 
\colhead{Luminosity\tablenotemark{b}}\\}
\startdata
3C\,454.3 & 86.12 & -38.1 & 2.053$\pm$0.02 & 2.47$\pm$0.01 & -54.7 & 
1.994$\pm$0.029 & 2.40$\pm$0.01 & 3.51$\pm$0.12 & 1.10$\pm$0.12 & 2.5$_{-
0.3}^{+0.3}$ & 0.859 & 4.8$_{-0.55}^{+0.55}$ & 44.1\\ 
PKS\,1502+106 & 11.37 & 54.58 & 1.068$\pm$0.02 & 2.24$\pm$0.01 & -34.0 & 
1.024$\pm$0.019 & 2.17$\pm$0.01 & 3.06$\pm$0.12 & 0.89$\pm$0.12 & 3.45$_{-
0.15}^{+0.9}$ & 1.839 & 9.7$_{-0.4}^{+2.6}$ & 185\\ 
3C\,279 & 305.1 & 57.06 & 0.754$\pm$0.01 & 2.32$\pm$0.02 & -6.60 & 0.724$\pm$0.021 
& 2.24$\pm$0.03 & 2.50$\pm$0.05 & 0.25$\pm$0.06 & 1.05$_{-0.2}^{+0.3}$ & 0.536 & 
1.6$_{-0.3}^{+0.5}$ & 5.0\\ 
PKS\,1510-08 & 351.2 & 40.13 & 0.739$\pm$0.02 & 2.47$\pm$0.02 & -7.13 & 
0.717$\pm$0.042 & 2.42$\pm$0.05 & 3.08$\pm$0.25 & 0.66$\pm$0.26 & 2.8$_{-0.6}^{+0.7}$ & 0.36 & 3.9$_{-0.8}^{+0.9}$ & 1.60\\ 
3C\,273 & 289.9 & 64.36 & 0.682$\pm$0.02 & 2.73$\pm$0.03 & -9.12 & 0.669$\pm$0.023 
& 2.68$\pm$0.03 & 3.66$\pm$0.28 & 0.97$\pm$0.28 & 1.6$_{-0.5}^{+0.5}$ & 0.158 & 1.9$_{-0.55}^{+0.55}$ & 0.19\\ 
PKS\,0454-234 & 223.7 & -34.9 & 0.632$\pm$0.01 & 2.19$\pm$0.01 & -23.5 & 0.604$\pm$0.016 & 2.11$\pm$0.02 & 3.28$\pm$0.21 & 1.16$\pm$0.21 & 4.95$_{-0.13}^{+1.1}$ & 1.003 & 9.95$_{-0.25}^{+2.1}$ & 22.1\\ 
PKS\,2022-07 & 36.89 & -24.3 & 0.439$\pm$0.01 & 2.38$\pm$0.03 & -4.54 & 
0.420$\pm$0.018 & 2.32$\pm$0.03 & 2.84$\pm$0.17 & 0.52$\pm$0.18 & 2.75$_{-0.18}^{+0.8}$ & 1.388 & 6.6$_{-0.45}^{+1.8}$ & 31.1\\ 
TXS\,1520+319 & 50.14 & 57.04 & 0.381$\pm$0.01 & 2.52$\pm$0.03 & -3.95 & 0.364$\pm$0.016 & 2.45$\pm$0.04 & 2.90$\pm$0.15 & 0.45$\pm$0.16 & 1.6$_{-0.4}^{+0.8}$ & 1.487 & 4.0$_{-0.1}^{+2.0}$ & 35.4\\ 
4C\,+38.41 & 61.08 & 42.40 & 0.229$\pm$0.01 & 2.50$\pm$0.04 & -10.1 & 
0.211$\pm$0.014 & 2.36$\pm$0.05 & 4.06$\pm$0.50 & 1.70$\pm$0.50 & 2.3$_{-
0.5}^{+0.55}$ & 1.814 & 6.4$_{-0.15}^{+1.5}$ & 39.3\\ 
PKS\,1908-201 & 16.89 & -13.1 & 0.160$\pm$0.01 & 2.42$\pm$0.06 & -9.31 & 
0.143$\pm$0.018 & 2.09$\pm$0.13 & 3.18$\pm$0.22 & 1.09$\pm$0.25 & 0.9$_{-0.25}^{+0.25}$ & 1.119 & 1.95$_{-0.55}^{+0.55}$ & 7.75\\ 
PKS\,0528+134 & 191.3 & -11.0 & 0.309$\pm$0.02 & 2.72$\pm$0.05 & -7.42 & 
0.268$\pm$0.020 & 2.52$\pm$0.07 & 3.87$\pm$0.47 & 1.34$\pm$0.48 & 1.6$_{-
0.4}^{+0.45}$ & 2.07 & 4.9$_{-1.2}^{+1.4}$ & 71.4\\ 
RGB\,J0920+446 & 175.7 & 44.82 & 0.247$\pm$0.01 & 2.22$\pm$0.03 & -7.61 & 
0.235$\pm$0.012 & 2.16$\pm$0.03 & 4.86$\pm$0.84 & 2.70$\pm$0.84 & 1.0$_{-0.5}^{+3.0}$ & 2.19 & 3.3$_{-1.5}^{+11.0}$ & 67.2\\ 
\enddata
\tablenotetext{a}{10$^{-6}$ ph[E$>$100 MeV]cm$^{-2}$s$^{-1}$}
\tablenotetext{b}{10$^{47}$ erg s$^{-1}$}
\end{deluxetable}

\begin{deluxetable}{cccccccccccccc}
\tablecolumns{12}
\tablewidth{0pc}
\tabletypesize{\tiny}
\tablecaption{Spectral properties of selected BLLacs}
\tablenum{2}
\rotate
\tablehead{
 \colhead{Name} & \colhead{l} & \colhead{b} & 
\colhead{Flux$_{PL}$\tablenotemark{a}} & \colhead{$\Gamma$} & \colhead{$\Delta 
L$} & \colhead{Flux$_{BPL}$\tablenotemark{a} } & \colhead{$\Gamma_1$} & 
\colhead{$\Gamma_2$}& \colhead{$\Delta \Gamma$} & \colhead{E$_{Break}$ (GeV)} & 
\colhead{z} & \colhead{E$\prime_{Break}$ (GeV)} & 
\colhead{Luminosity\tablenotemark{b}}\\}
\startdata
\sidehead{LSP-BLLacs}
AO\,0235+164 & 156.7 & -39.0 & 0.630$\pm$0.01 & 2.12$\pm$0.01 & -20.8 & 
0.599$\pm$0.016 & 2.04$\pm$0.02 & 2.80$\pm$0.12 & 0.75$\pm$0.13 & 4.5$_{-1.0}^{+1.5}$ & 0.94 & 8.8$_{-1.9}^{+3.0}$ & 19.6\\ 
PKS\,0537-441 & 250.0 & -31.0 & 0.400$\pm$0.01 & 2.28$\pm$0.02 & -8.51 & 
0.380$\pm$0.015 & 2.20$\pm$0.03 & 3.08$\pm$0.23 & 0.87$\pm$0.24 & 3.8$_{-
0.8}^{+1.3}$ & 0.892 & 7.1$_{-1.5}^{+2.5}$ & 10.1\\ 
PKS\,0426-380 & 240.7 & -43.6 & 0.274$\pm$0.01 & 2.18$\pm$0.03 & -7.17 & 
0.255$\pm$0.012 & 2.10$\pm$0.03 & 3.36$\pm$0.45 & 1.26$\pm$0.45 & 8.3$_{-0.4}^{+6.0}$ & 1.112 & 17.5$_{-0.8}^{+13.0}$ & 12.3\\ 
\sidehead{HSP-BLLac}
1ES\,0502+675 & 143.7 & 15.89 & 0.019$\pm$0.00 & 1.70$\pm$0.14 & -8.40 & 0.064$\pm$0.015 & 2.68$\pm$0.18 & 1.47$\pm$0.10 & -1.2$\pm$0.21 & 1.4$_{-0.5}^{+0.7}$ & 0.416 & 2.0$_{-0.7}^{+1.0}$ & 0.12\\ 
\enddata
\tablenotetext{a}{10$^{-6}$ ph[E$>$100 MeV]cm$^{-2}$s$^{-1}$}
\tablenotetext{b}{10$^{47}$ erg s$^{-1}$}
\end{deluxetable}

\end{document}